\documentclass[onecolumn,nofootinbib,11pt]{revtex4-1}

\usepackage{graphicx}
\usepackage{epstopdf}
\usepackage{latexsym}
\usepackage{amsmath}
\usepackage{amssymb}
\usepackage{color}
\usepackage{mathrsfs}

\usepackage[center]{subfigure}

\begin{document}

 \newcommand{\bq}{\begin{equation}}
 \newcommand{\eq}{\end{equation}}
 \newcommand{\bqn}{\begin{eqnarray}}
 \newcommand{\eqn}{\end{eqnarray}}
 \newcommand{\nb}{\nonumber}
 \newcommand{\lb}{\label}
 \newcommand{\be}{\begin{equation}}
\newcommand{\en}{\end{equation}}
\newcommand{\PRL}{Phys. Rev. Lett.}
\newcommand{\PL}{Phys. Lett.}
\newcommand{\PR}{Phys. Rev.}
\newcommand{\CQG}{Class. Quantum Grav.}

\title{Echoes of Stars in tracking Rastall gravity}

\author{Kai Lin $^{a}$}\email{lk314159@hotmail.com}
\author{Chieh-Hung Chen$^{a}$}\email{nononochchen@gmail.com}
\author{Yang-Yi Sun $^{a}$}\email{sunyy@cug.edu.cn}
\author{Xilong Fan$^{b}$}\email{xilong.fan@whu.edu.cn}
\author{Hongsheng Zhang$^{c}$}\email{sps_zhanghs@ujn.edu.cn (corresponding author)}

\affiliation{a) School of Geophysics and Geomatics, China University of Geosciences, Wuhan 430074, Hubei, China}
\affiliation{b) School of Physics and Technology, Wuhan University, Wuhan 430072, China}
\affiliation{c) School of Physics and Technology, University of Jinan, 336 West Road of Nan Xinzhuang, Jinan 250022, China}

\date{\today}

\begin{abstract}
 We construct a star with uniform density in frame of tracking Rastall gravity, in which the density of effective dark matter tracks the density of dark energy. To
 demonstrate stability of the star we analyse the axial gravitational perturbations of the star. We find that echo wavelets naturally appear in the axial gravitational waves
 when the dark matter plays a significant role in construction of the star. We show that the physical origin of this echo roots in the potential valley near the surface of
 star. This effect leads to a clue to detect Rastall dark energy from the ringdown phase of gravitational wave from binary stars enriched by dark matter.
\end{abstract}

\pacs{04.60.-m; 98.80.Cq; 98.80.-k; 98.80.Bp}

\maketitle

\section{Introduction}
\renewcommand{\theequation}{1.\arabic{equation}} \setcounter{equation}{0}

In 1972\cite{Rastall}, Rastall questioned the validity of the conservation law of the stress-energy in curved spacetime. He argued that the relation $\partial_\mu T^\mu_\nu=0$
is valid in flat spacetime, but there is no evidence and observation to verify that the relation can be directly generalized in curved spacetime by rewriting ordinary
derivative as covariant derivative. Therefore, a reasonable assumption is the covariant derivative of the stress-energy tensor does not vanish. Then, Rastall introduces a
simple term as follows,
\bqn
\lb{RastallA}
\nabla_\mu T^\mu_\nu=\lambda\nabla_\nu R
\eqn
so that the gravitational field equation becomes
\bqn
\lb{RastallB}
R_{\mu\nu}-\frac{1}{2}g_{\mu\nu}R&=&8\pi G\left(T_{\mu\nu}-\lambda g_{\mu\nu}R\right).
\eqn
In fact, the original gravity theory of Einstein and Grossmann can be treated as a special case with $\lambda=1/4\pi G$ \cite{EG}. In this sense, Rastall gravity is a direct
extension of gravity theory of Einstein and Grossmann presented in 1913.
Recently, Rastall gravity gets a lot of attention in astrophysics and theoretical physics. Several Rastall black hole and star solutions are obtained in
\cite{RastallT1,RastallT2,RastallT3,RastallT4,RastallT5,RastallT6,RastallT7,RastallT8,RastallT9,RastallT10,Lin1,Lin2}, and lots of cosmological effect are discussed in
\cite{RastallT11,RastallT12,RastallT13,RastallT14,RastallT15,RastallT16,RastallT17,RastallT18,RastallT19,RastallT20,RastallT21,Lin3}.
The Rastall gravity is consistent with recent astronomical observation \cite{RastallT22}.

Inspired by the original idea of Rastall, we propose a generalized Rastall gravity \cite{Lin1,Lin2,Lin3} which gravitational field and the modified conservation law of the
stress energy-momentum tensor satisfy,
\bqn
\lb{FieldEqu1}
R_{\mu\nu}-\frac{1}{2}g_{\mu\nu}R&=&8\pi G\left(T_{\mu\nu}-{\cal A}_{\mu\nu}\right)\nb\\
\nabla_\mu T^\mu_\nu&=&\nabla_\mu {\cal A}^\mu_\nu
\eqn
where the Rastall theory requires that ${\cal A}_{\mu\nu}$ must vanish at flat spacetime. This theory greatly expands the range of application of the Rastall gravity. Many
modified gravitational theories are equivalent to Rastall gravity. Physically ${\cal A}_{\mu\nu}$ denotes matter creation/annihilation, which is balanced by a non-conserved
$T_{\mu\nu}$. In this sense it revives an early theory of steady state Universe of Hoyle \cite{FH} but with a generalized form of matter creation tensor.

The essential feature of Rastalll gravity is that energy flows from ${\cal A}_{\mu\nu}$ to $T_{\mu\nu}$ or reversely. If $T_{\mu\nu}$ include dark matter and ${\cal
A}_{\mu\nu}$ represent dark energy, the Rastall gravity yields an interacting dark energy model. Such models have been widely investigated in cosmology \cite{zhang}.  In
\cite{Lin3}, based on the generalized Rastall gravity, a new dark energy model without cosmic coincidence paradox is proposed. In this model we command the density of dark
energy always to track the density of dark matter, in analogy to the tracking solution of quintessence model \cite{SWZ}.  Here we name the model in \cite{Lin3} tracking
Rastall gravity. In the model, the Rastall term is given by
\bqn
\lb{Rastall1}
{\cal A}_{\mu\nu}=g_{\mu\nu}\rho_\text{de}=\beta g_{\mu\nu}\rho_\text{dm}
\eqn
where $\rho_\text{de}$ and $\rho_\text{dm}$ are the densities of dark energy and dark matter respectively. $\beta\equiv\frac{\rho_\text{de}}{\rho_\text{dm}}$ is a constant, so
that the ratio between dark energy and dark matter remains unchanged, and the cosmic coincidence paradox is avoided.

Different from the standard $\Lambda$CDM model, the effect of ordinary matter $\rho_m$ and dark matter $\rho_\text{dm}$ are different in the tracking Rastall cosmological
model, because dark energy and dark matter interact each other. If a star in the model includes a component of dark matter, the star must be affected by dark energy. It is
interesting to discuss the Rastall dark energy effect in the star spacetime.   We derive an analytical solution for star in spherically symmetry within the tracking Rastall
gravity and its axial gravitational perturbations. We find that the star with high proportioned dark matter produces axial gravitational wave with echo wavelet.

 This article is organized as follows. In the next section, we present the exact solution of stars with uniform density in tracking Rastall gravity. In section III, we study
 axial gravitational perturbation, and find that the potential develops a valley when dark matter takes a significant role in the total density. By aids of this valley the
 echoes follow the main wave in merging process of binary stars. Section IV concludes this paper.

\section{The analytical Solution of the relativistic star with Rastall dark energy}
\renewcommand{\theequation}{2.\arabic{equation}} \setcounter{equation}{0}

In order to construct the spherically symmetric relativistic star solution, we choose the ansatz as the static metric
\bqn
\lb{metric1}
ds^2=-f(r)dt^2+\frac{dr^2}{h(r)}+r^2\left(d\theta^2+\sin^2\theta d\varphi^2\right),
\eqn
and the stress-energy tensor inside of the star is considered to be constituted by the prefect fluid matter and dark matter, namely
\bqn
\lb{EnergyTensor1}
T^\mu_\nu&=&(\rho_m+\rho_\text{dm}+P)U_\nu U^\mu+P\delta^\mu_\nu\nb\\
&=&\text{diag}\left[-\rho_m(r)-\rho_\text{dm}(r),P(r),P(r),P(r)\right]
\eqn
where $U_\mu=\delta^t_\mu\sqrt{f(r)}$ and $P(r)$ is intensity of pressure inside of the star. Substituting above stress-energy tensor and metric into Rastall gravitational
field equation, we can get 3 independent equations
\bqn
\lb{FieldEqu2}
h'&=&\frac{1-h}{r}-8\pi G r\left[\rho_m+(1-\beta)\rho_\text{dm}\right],\nb\\
f'&=&\frac{f}{rh}\left[1-h+8\pi G r^2 (P+\beta\rho_\text{dm})\right],\nb\\
P'&=&\frac{(P+\rho_m+\rho_\text{dm})}{2rh}\left[h-1-8G\pi r^2\left(P+\beta\rho_\text{dm}\right)\right]\nb\\
&&-\beta\rho'_\text{dm}.
\eqn
It is easy to find that the 3 independent equations include 5 undetermined functions $h(r)$, $f(r)$, $\rho_m$, $\rho_\text{dm}$ and $P(r)$, so it implies that we need the
other 2 independent state equations $\rho_m=\rho_m(P,\rho_\text{dm})$ and $\rho_\text{dm}=\rho_\text{dm}(P,\rho_m)$ to completely solve above functions.

On the other hand, the spacetime outside of the star can be described by Schwarzschild solution, which is written as
\bqn
\lb{FieldEqu3}
h_E=f_E=1-\frac{2GM}{r},\nb\\
P_E=\rho_\text{mE}=\rho_\text{dmE}=0
\eqn
where the index $I$ and $E$ represent interior solution and exterior solution respectively, and the functions at surface of star $r=r_s$ should satisfy the connect condition
\bqn
\lb{FieldEqu4}
h_I(r_s)&=&h_E(r_s),  ~~~f_I(r_s)=f_E(r_s),\nb\\
P_I(r_s)&=&0
\eqn

Generally, it is very hard to analytically solve about equations (\ref{FieldEqu2}). However, we can still obtain the simplest analytical solution, which requires $\rho_m$ and
$\rho_\text{dm}$ are constants, and the relativistic star becomes a star of uniform density:
\bqn
\lb{FieldEqu5}
h_I(r)&=&1-\frac{8}{3}G\pi r^2\rho_o=1-\frac{2GM_I(r)}{r},\nb\\
f_I(r)&=&\frac{\left[3\rho_A\sqrt{h_I(r_s)}-\rho_B\sqrt{h_I(r)}\right]^2}{4\rho_o^2},\nb\\
P_I(r)&=&\frac{\rho_A\rho_B\left(\sqrt{h_I(r)}-\sqrt{h_I(r_s)}\right)}{3\rho_A\sqrt{h_I(r_s)}-\rho_B\sqrt{h_I(r)}},
\eqn
where $\rho_o=\rho_m+(1-\beta)\rho_\text{dm}$, $\rho_A=\rho_o+\beta\rho_\text{dm}$, $\rho_B=\rho_o+3\beta\rho_\text{dm}$ and $M_I(r)=\frac{4}{3}\pi r^3$ which satisfy the
connect condition $M_I(r_s)=M$ as the effect total mass.

To further simplify the solution, we can introduce the dimensionless physical quantities
\bqn
\lb{FieldEqu6}
x&=&\frac{r}{r_s},~~~~~\rho_\text{dm}=\alpha\rho_o,,\nb\\
M_\text{eff}&=&G\frac{M}{r_s},~~~V_s=\frac{4}{3}G\pi r_s^2.
\eqn
so the solution becomes
\bqn
\lb{FieldEqu7}
\hat{h}_I(x)&=&1-2M_\text{eff}x^2,\nb\\
\hat{f}_I(x)&=&\frac{1}{4}\left[3(1+\alpha\beta)\sqrt{\hat{h}_I(x)}-(1+3\alpha\beta)\sqrt{\hat{h}_I(1)}\right]^2,\nb\\
\hat{P}_I(x)&=&\frac{M_\text{eff}}{V_s}\frac{(1+\alpha\beta)(1+3\alpha\beta)(\sqrt{\hat{h}_I(1)}-\sqrt{\hat{h}_I(x)})}{(1+3\alpha\beta)\sqrt{\hat{h}_I(x)}-3(1+\alpha\beta)\sqrt{\hat{h}_I(1)}},\nb\\
\hat{f}_E(x)&=&\hat{h}_E(x)=1-\frac{2M_\text{eff}}{x}.
\eqn

Let's fix the dimensionless effect total mass $M_\text{eff}=0.3$, and show the functions $\hat{f}$, $\hat{h}$ and $\hat{P}$ in FIG.(1) as follows
\begin{figure*}[tbp]
\lb{FigA}
\centering
\includegraphics[width=0.45\columnwidth]{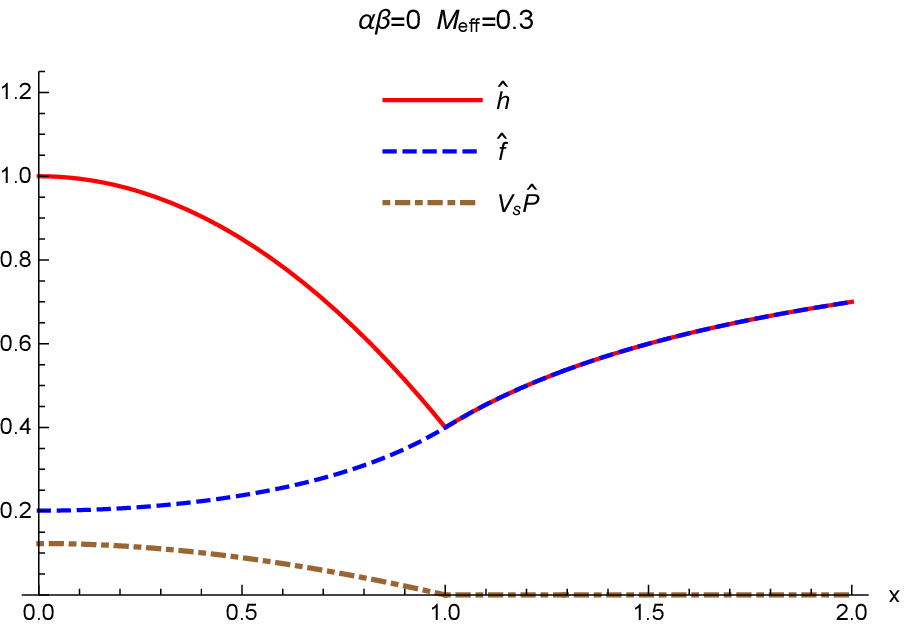}\includegraphics[width=0.45\columnwidth]{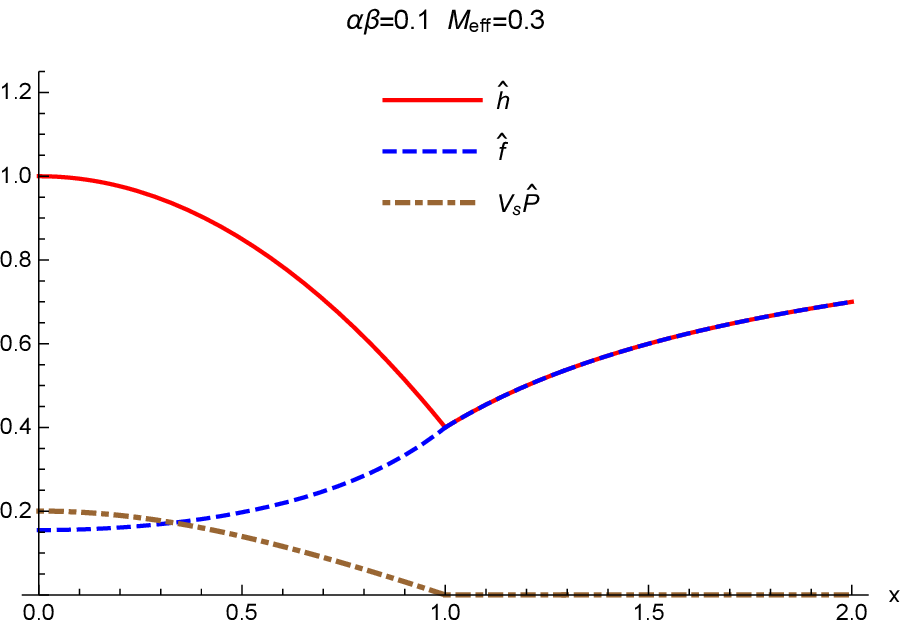}
\includegraphics[width=0.45\columnwidth]{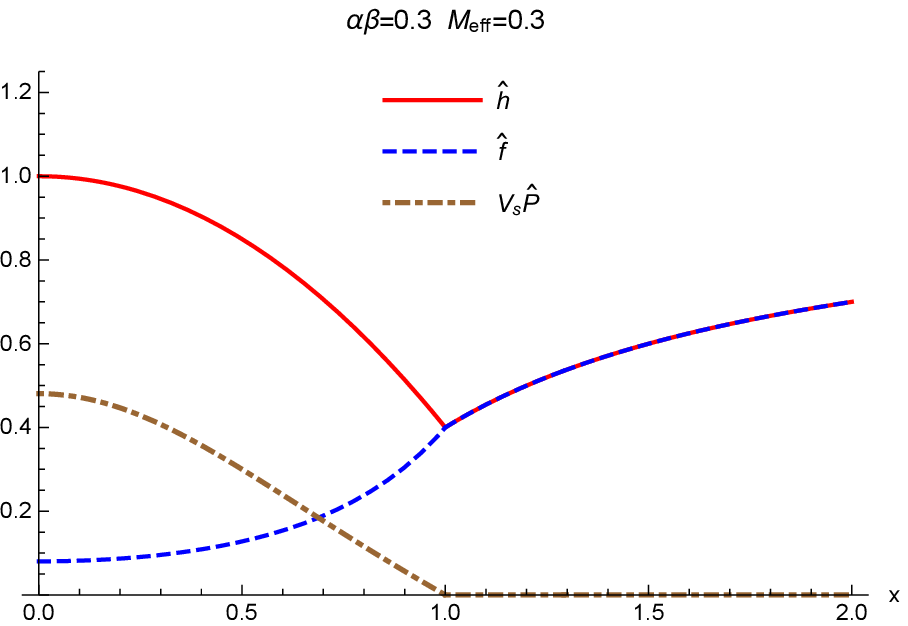}\includegraphics[width=0.45\columnwidth]{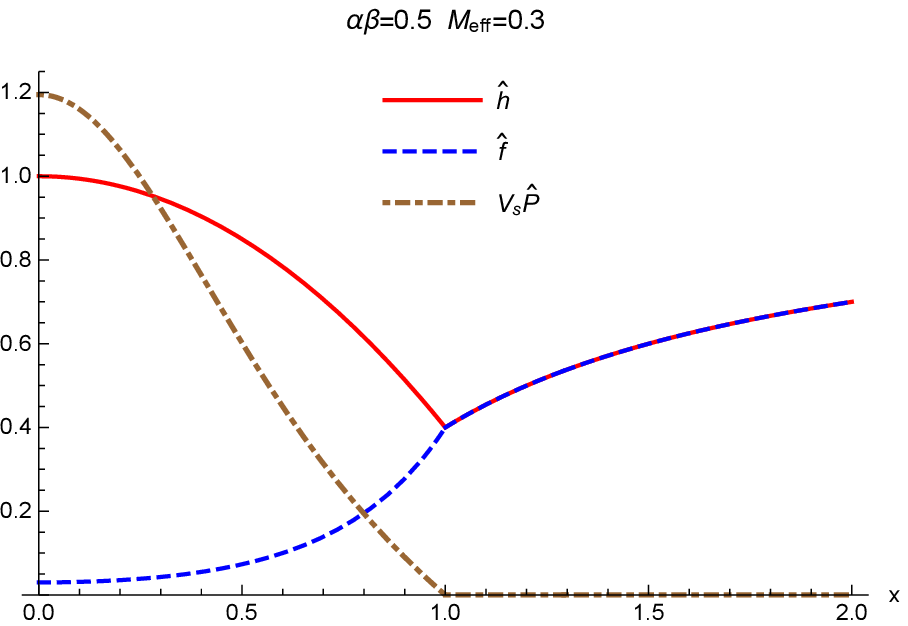}
\caption{The Relativistic Star Solutions of Uniform Density}
\end{figure*}

From above Figure and Eq.(\ref{FieldEqu7}), we find the function $\hat{h}(x)$ only depends on parameter $M_\text{eff}$, but the functions $\hat{f}(x)$ and $\hat{P}(x)$ depend
on both of parameters $M_\text{eff}$ and $\alpha\beta$. Especially, the intensity of pressure $\hat{P}(x)$ increases rapidly as higher $\alpha\beta$, and it is thought as the
effect of Rastall dark energy which can provide pressure for universe accelerating expanding in cosmology. Next section, we will study axial gravitational perturbation in the
spacetime, and emphatically discuss the effect from large value of parameter $\alpha\beta$.

\section{axial gravitational perturbation and its dynamical evolution}
\renewcommand{\theequation}{3.\arabic{equation}} \setcounter{equation}{0}

In 2015, Laser Interferometer Gravitational Wave Observator (LIGO) successfully detected a gravitational wave (GW) signal GW150914 from binary black hole merger\cite{LIGO1},
and then more and more GW signals are found\cite{LIGO2,LIGO3,LIGO4,LIGO5}. According to the waveform of GW, the binary black hole GW includes 3 stage: Inspiral, Merge and
Ringdown. Ringdown phase is from the final stage of binary black hole merger. At this time, two black holes have been merging into one, but the final black hole is still
oscillating, and the Ringdown GW is produced by the eigenvibration of black hole. Therefore, we can study this stage by the perturbation and Quasinormal modes theory of black
hole. Because Quasinormal modes and Ringdown GW includes abundant information of black hole, which will help to understand theoretical physics and astrophysics, the research
about Ringdown get a lot of attention of scientists.

Soon after the GW150914, the Advanced LIGO and Virgo detectors get the GW signal from the binary neutron star merge\cite{LIGOStar1,LIGOStar2}, so that it marks the beginning
of the era of multimessenger astronomy. Ringdown phase of binary neutron star's merge playes a more and more important role in the research of astronomy. However, the
dynamical evolution of compact star's ringdown GW are more hard to be studied than black hole's ringdown GW, because we must calculate the perturbation inside of the star and
the outside part at the same time, and then connect the results at the surface of the neutron star.

In order to calculate the axial gravitational perturbation, let's choose Regge-Wheeler gauge \cite{ReggeWheeler} which are given by
\bqn
\lb{metric3}
\delta g_{\mu\nu}&=&
\begin{bmatrix}
0 & 0 & 0 & h_0(r) \\
0 & 0 & 0 & h_1(r) \\
0 & 0 & 0 & 0  \\
h_0(r) & h_1(r) & 0 & 0
\end{bmatrix}
e^{-i\omega t}\sin\theta\partial_\theta P_L(\cos\theta)\nb\\
\delta u_\mu&=&\delta^\varphi_\mu h_n(r)e^{-i\omega t}\sin\theta\partial_\theta P_L(\cos\theta)\nb\\
\delta U_\mu&=&\delta^\varphi_\mu h_w(r)e^{-i\omega t}\sin^{-1}\theta\partial_\theta P_L(\cos\theta)
\eqn
where the perturbation is much small than the background metric, namely $g_{\mu\nu}\gg\delta g_{\mu\nu}$, $u_{\mu}\gg\delta u_{\mu}$ and $U_{\mu}\gg\delta U_{\mu}$.

Finally, we can get the axial gravitational perturbation equation
\bqn
\lb{PertEqu1}
\sqrt{fh}\frac{\partial}{\partial r}\left[\sqrt{fh}\frac{\partial \Psi}{\partial r}\right]+\left(\omega^2-V(r)\right)\Psi=0
\eqn
where $V$ is the potential function of perturbation equation, and the function form inside and outside of the relativistic star respectively are given by
\bqn
\lb{PertEqu2}
V_I(r)&=&\frac{f}{r^2}\left[L^2+L-3\right.\nb\\
&&\left.+4\pi Gr^2\left(\rho_m+(1-2\beta)\rho_\text{dm}-P\right)\right],\nb\\
V_E(r)&=&\frac{f}{r^2}\left[L^2+L-\frac{6M}{r}\right].
\eqn

Let's rewrite above perturbation equation by the dimensionless physical quantities, so we obtain
\bqn
\lb{PertEqu3}
&&\sqrt{\hat{f}_I\hat{h}_I}\frac{\partial}{\partial x}\left[\sqrt{\hat{f}_I\hat{h}_I}\frac{\partial \Psi_I}{\partial
x}\right]+\left[w^2-\frac{\hat{f}_I}{x^2}\left(L^2+L\right.\right.\nb\\
&&\left.\left.-3+3M_\text{eff}(1-\alpha\beta)x^2+3\hat{h}_I-3\hat{P}_Ix^2\right)\right]\Psi_I=0\nb\\
&&\hat{f}_E\frac{\partial}{\partial x}\left[\hat{f}_E\frac{\partial \Psi_E}{\partial x}\right]+\left[w^2-\frac{\hat{f}_E}{x^2}\left[L^2+L\right.\right.\nb\\
&&\left.\left.-3+3\hat{h}_E\right)\right]\Psi_E=0
\eqn
where $w=\omega r_s$ is dimensionless frequency.

The boundary condition of relativistic star spacetime requires that there is only outgoing model at infinity, but the wave function keeps finite at center of the star, namely
\cite{Connect}
\bqn
\lb{PertEqu4}
\Psi_I&\sim& x^{L+1}~~~~~~~~~~~~~~~~~~~~~~~~~\text{at center}~x=0\nb\\
\Psi_E&\sim& e^{iwx}\left(x-2M_\text{eff}\right)^{2iM_\text{eff}w}~\text{at infinity}~x\rightarrow\infty\nb\\
\eqn

What's more the connect condition \cite{Connect} at the surface of star requires are given by
\bqn
\lb{PertEqu5}
\frac{\Psi_I'(1)}{\Psi_I(1)}=\frac{\Psi_E'(1)}{\Psi_E(1)}~~~~~~~~~~~~~~\text{at surface}~x=1
\eqn
In order to study the dynamical evolution of axial gravitational perturbation in the relativistic star spacetime, we use the transformation $w\rightarrow i\partial_t$ to
rewritten frequency domain perturbation equation (\ref{PertEqu3}) as time domain perturbation equation, and then introduce the tortoise coordinate transformation
$x_*=\int_{0}^{x} {dx'}/{\sqrt{\hat{f}(x')\hat{h}(x')}}$, so that
\bqn
\lb{PertEqu7}
\frac{\partial^2\Psi}{\partial x_*^2}-\frac{\partial^2\Psi}{\partial t^2}-V(x)\Psi=0
\eqn

The potential function $V(x)$ here is shown in FIG.2.
\begin{figure}[tbp]
\centering
\includegraphics[width=0.8\columnwidth]{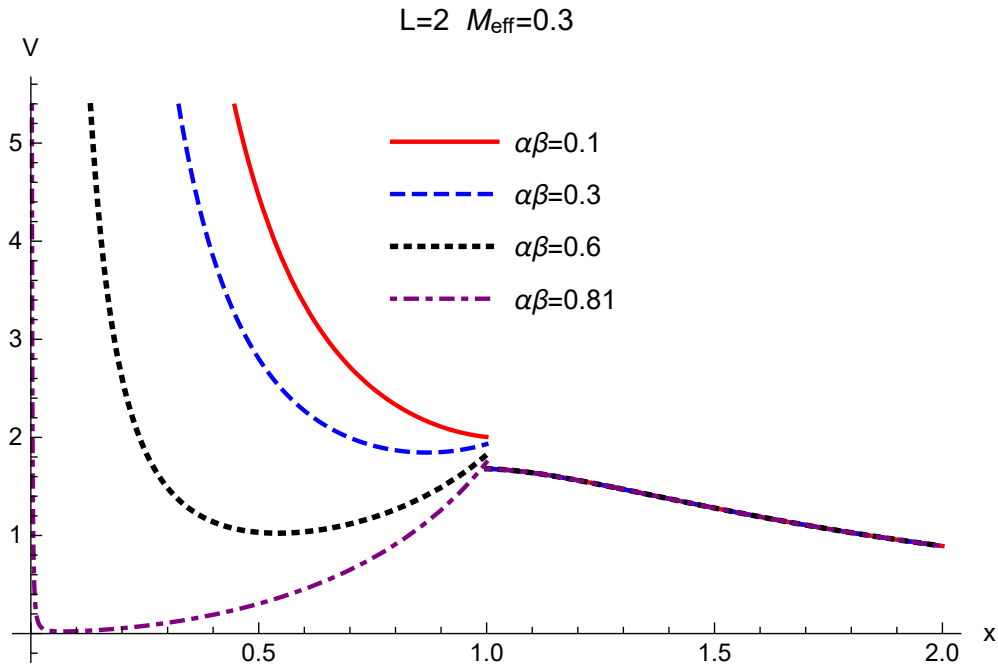}
\caption{The Potential $V$ with different parameters}
\lb{Fig2}
\end{figure}
It is interesting there is a potential valley inside  the star as high proportion of dark matter in the star, which could produce echo wavelets in gravitational wave
signals\cite{echo}.

By using finite difference method (which details is shown in \cite{Lin4}), we can get the process of dynamical evolution of axial gravitational perturbation, which is shown in
Fig.3.
\begin{figure*}[tbp]
\centering
\includegraphics[width=0.45\columnwidth]{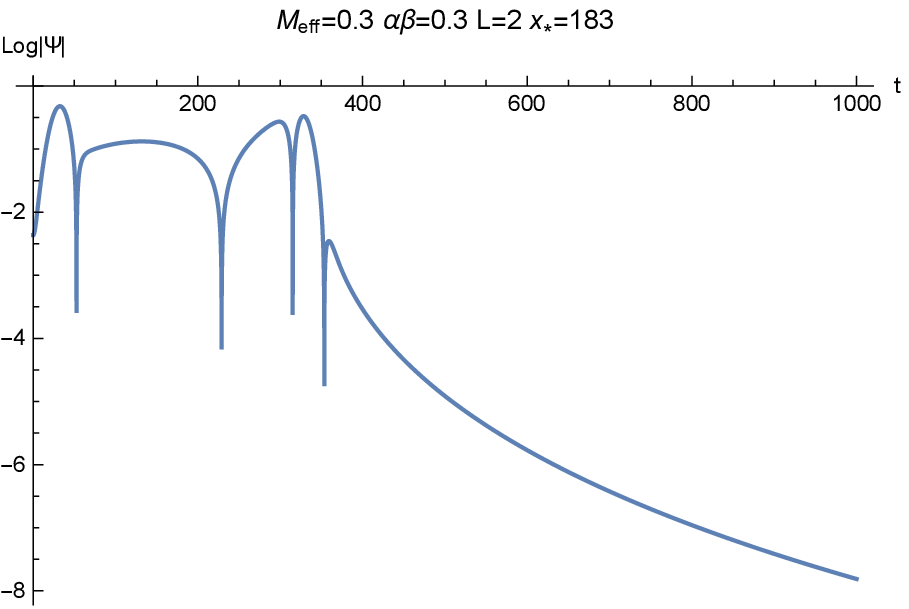}\includegraphics[width=0.45\columnwidth]{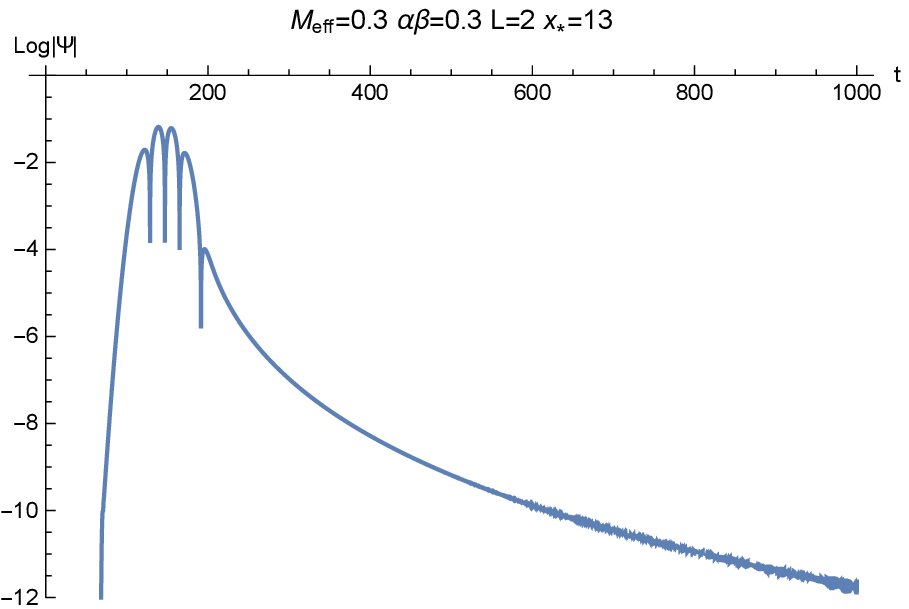}
\includegraphics[width=0.45\columnwidth]{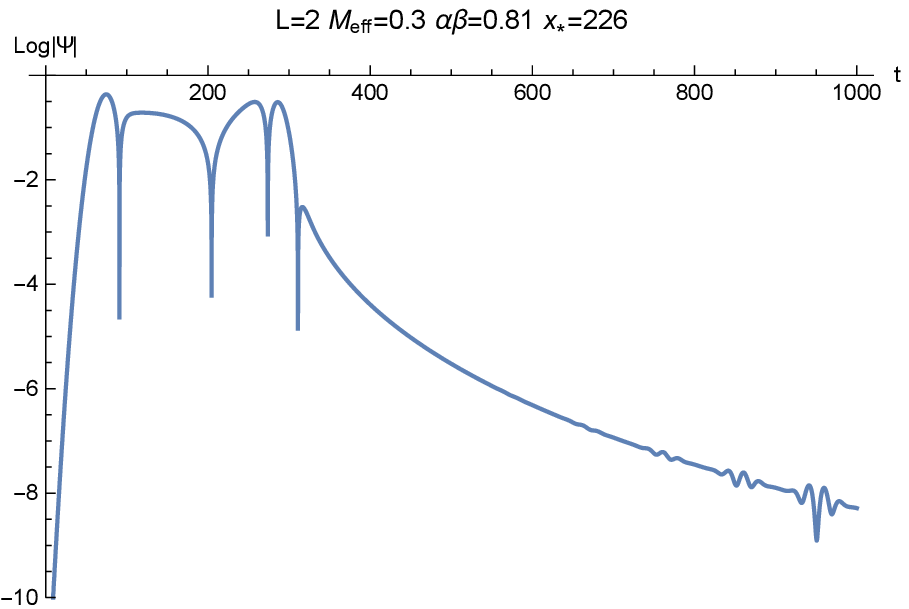}\includegraphics[width=0.45\columnwidth]{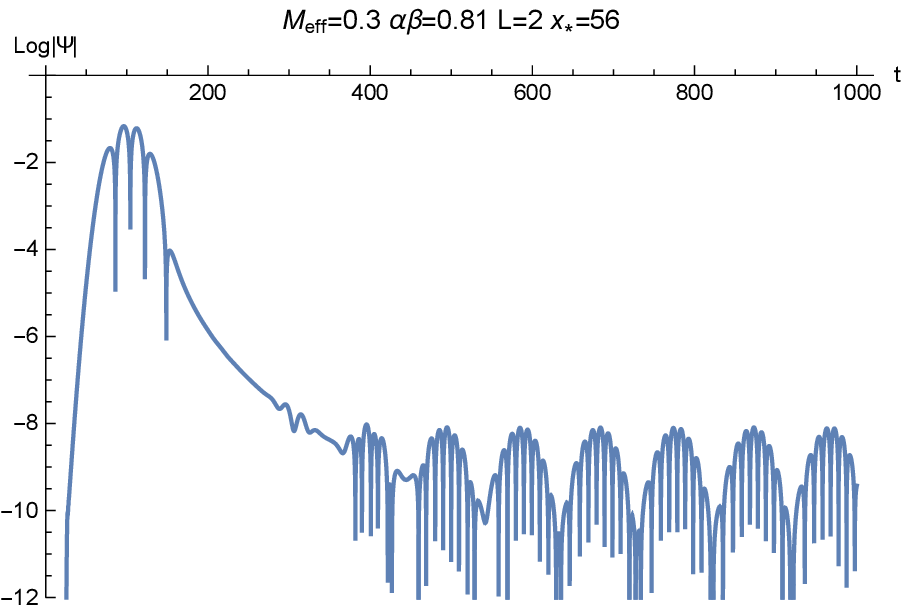}
\caption{Dynamical Evolution of Axial Gravitational Perturbation}
\lb{Fig3}
\end{figure*}

Above figure shows the axial gravitational perturbation, and we find the echo wavelet appears as $\alpha\beta=0.81$, but there is only power-law tail as $\alpha\beta=0.3$. The
phenomena provide us a valid method to detect Rastall dark energy and dark matter in relativistic star.

\section{Conclusion}
\renewcommand{\theequation}{4.\arabic{equation}} \setcounter{equation}{0}

There are several motivations to consider a theory with non-conserved stress energy. The
kinetic diffusive process, which is described by Fokker-Planck equation, leads to a non-conserved stress energy \cite{non1}. Quantum effects in curved space yield
non-conserved
 stress energy either \cite{non2}. A straightforward example is a black hole in box. At first the hole is in vacuum and stress energy is zero. After a period of time,
 radiations generated by Hawking evaporation appear in the box. Clearly, classical stress energy is not conserved. Rastall proposed a theory without conservation of classical
 stress energy. We extend it to a more generic form to describe the tracking behaviour of dark matter and dark energy.

In this paper, we studied the relativistic star with Rastall dark energy and its axial gravitational perturbation in frame of tracking Rastall gravity. We obtain an analytical
solution of the star with uniform density ordinary matter and dark matter. Because of tracking behaviour of dark energy and dark matter in this model, the inside structure of
the star is strongly influenced by Rastall dark energy with negative pressure, but the outside spacetime of the star is Schwarzschild vacuum spacetime because the dark matter
vanishes there. What's more, the effect of dark energy leads to a potential valley inside of the star, so that there is echo wavelet in axial GW from the relativistic star
spacetime. The property is helpful to be detected in GW signals.

\begin{acknowledgments}
We gratefully acknowledge the financial support from
National Natural Science Foundation of China (NNSFC) under contract Nos. 42230207, 12275106, 12235019, 11922303 and 11805166.
\end{acknowledgments}

\end{document}